\documentclass{elsart}
\date{}

\usepackage[latin1]{inputenc}
\usepackage{amssymb}

\newtheorem{The}{Theorem}[section]
\newtheorem{Pro}[The]{Proposition}
\newtheorem{Deff}[The]{Definition}
\newtheorem{Lem}[The]{Lemma}
\newtheorem{Rem}[The]{Remark}
\newtheorem{Cor}[The]{Corollary}

\newcommand{\fa}{\forall}
\newcommand{\Ga}{\Gamma}
\newcommand{\Gas}{\Gamma^\star}
\newcommand{\Si}{\Sigma}
\newcommand{\Sis}{\Sigma^\star}
\newcommand{\ra}{\rightarrow}

\newcommand{\lra}{\leftrightarrow}
\newcommand{\la}{language}
\newcommand{\ite}{\item}

\newcommand{\ol}{ $\omega$-language}
\newcommand{\orl}{ $\omega$-regular language}
\newcommand{\om}{\omega}
\newcommand{\nl}{\newline}
\newcommand{\noi}{\noindent}
\newcommand{\tla}{\twoheadleftarrow}
\newcommand{\de}{deterministic }

\newcommand{\proo}{\noi {\bf Proof.} }
\newcommand {\ep}{\hfill $\square$}

\begin{document}
\begin{frontmatter}
\title{\bf BOREL HIERARCHY AND OMEGA CONTEXT FREE LANGUAGES}

\author{Olivier Finkel\corauthref{cor}}
\corauth[cor]{Corresponding author}
\ead{finkel@logique.jussieu.fr }

\address{ Equipe de Logique Math\'ematique \\CNRS  et 
Universit\'e Paris 7,  U.F.R. de Math\'ematiques \\  2 Place Jussieu 75251 Paris
 cedex 05, France.}

\begin{abstract}
\noi  We give in this paper additional answers to questions of 
Lescow and  Thomas [Logical Specifications of Infinite Computations,
 In:"A Decade of Concurrency", 
Springer LNCS 803 (1994), 583-621], proving  topological 
properties of  omega context free  languages 
($\om$-CFL) which extend those of 
[O. Finkel, Topological Properties of Omega Context Free Languages, 
 Theoretical Computer Science, Vol. 262 (1-2), 2001, p. 669-697]: there exist some 
$\om$-CFL which are non Borel sets and one cannot decide whether an $\om$-CFL is a Borel set.
We give also an answer to a question of  
 Niwinski [Problem on $\omega$-Powers Posed  in the Proceedings 
of the 1990 Workshop "Logics and Recognizable Sets"] and  of 
Simonnet [Automates et Th\'eorie Descriptive, Ph.D. Thesis, Universit\'e 
Paris 7, March 1992] about $\om$-powers of finitary 
languages, giving an example of a finitary context free language $L$ such that $L^\om$ 
is not a Borel set.
 Then we prove some recursive analogues to preceding  properties: in particular one 
cannot decide whether an $\om$-CFL is an arithmetical set. Finally we extend some 
results to context free sets of infinite trees.
\end{abstract}

\end{frontmatter}

\section{Introduction}

Since  B\"uchi studied the \ol s recognized   by finite   automata to 
prove 
the decidability of the monadic second order theory of one successor
 over the integers \cite{bu60a}
 the so called $\om$-regular languages have been intensively
 studied.
See \cite{tho} and \cite{pp} for many results and references.

As pushdown automata are a natural extension of finite automata, 
 Cohen and  Gold \cite{cg} , \cite{cgdet} 
  and  Linna \cite{lin76}  studied the \ol s accepted by omega 
pushdown automata,
 considering various acceptance conditions for omega words.
It turned out that the omega languages accepted by  omega pushdown 
automata were also those
 generated by context free grammars where infinite derivations are
 considered , also studied by 
 Nivat \cite{ni77}, \cite{ni78}  and  Boasson and  Nivat \cite{bn}.  These languages were then 
called the 
 omega context free  languages ($\om$-CFL).
See also Staiger's paper \cite{sta} for a survey of general theory of
\ol s.

Topological properties of \orl s were first
 studied by  Landweber in \cite{la} where he showed that these \la s are boolean 
combinations of $G_\delta$ sets. He also characterized the \orl s in each of the 
Borel classes 
 ${\bf F, G, F_\sigma, G_\delta }$, and showed that one can decide, for an effectively given
\orl~ $L$, whether $L$ is in the Borel class  ${\bf F, G, F_\sigma}$, or  ${\bf G_\delta }$.
\nl It turned out that an \orl~ is in the class ${\bf G_\delta }$  iff it is accepted 
by a \de B\"uchi automaton.
These results were extended to \de $\om$-CFL by Linna \cite{lin77}. 
 In the non \de case,  Cohen and Gold proved in \cite{cgdet}
 that one cannot decide whether an $\om$-CFL is in the class 
${\bf F, G }$ or ${\bf G_\delta}$.

   We have begun a similar study for $\om$-CFL in \cite{fina}.
 we proved that $\om$-CFL exhaust the finite ranks 
 of the  Borel hierarchy and  that, for any Borel class ${\bf \Si_n^0 }$ or
 ${\bf \Pi_n^0}$, n being an  integer, one cannot decide whether an $\om$-CFL is 
in  ${\bf \Si_n^0}$ or ${\bf \Pi_n^0}$. 
Our proof used the Wadge game and the operation of exponentiation 
of sets defined by  Duparc \cite{dup}.

   We pursue this study  in this paper. We first show that there exist some  
$\om$-CFL which are analytic but non Borel sets. 
Then we extend the preceding undecidability result to every Borel class 
(of finite {\it or} infinite rank) and we prove that one cannot even decide 
whether an $\om$-CFL is a Borel set. 

   The question of the topological complexity of the $\om$-power of a finitary 
language is mentioned in \cite{sta} \cite{staop}. Niwinski asked  in \cite{niw} 
for an example of a (finitary) language $L$ such that $L^\om$ is not a Borel set. 
Simonnet asked in \cite{sim} for the topological complexity of $L^\om$ where $L$
 is a context free language.  We proved in \cite{fina} that there exist 
context free languages $L_n$ such that $(L_n)^\om$ is a ${\bf \Pi_n^0}$-complete set 
for each integer $n\geq 1$. 

    We give here an example of a context free 
language $L$ such that $L^\om$ is an analytic but not Borel set, answering to questions of 
Niwinski and Simonnet.

   Then  we derive  some new arithmetical properties of omega context free \la s 
from the preceding topological properties. We prove that one cannot decide whether an 
$\om$-CFL is 
 an arithmetical set in $\bigcup_{i\geq 1}\Si_n$.
Then we show that one cannot decide whether the complement of  an $\om$-CFL is 
accepted by a (non deterministic) 
Turing machine (or more generally by a non \de {\bf X}-automaton as defined in \cite{eh}) with  
B\"uchi (respectively 
Muller) acceptance 
condition.
 The above  results  give additional answers to questions 
of Thomas and  Lescow \cite{lt}.

 Finally we extend some undecidability results to context free sets of infinite trees,
as defined by Saoudi \cite{sao}.

 The paper is organized as follows.
In sections 2 and 3, we first review some above definitions and results about $\om$-regular,
 $\om$-context free languages, and topology.
Then in section 4 we prove our main topological results from which  we 
deduce in section 5 the result 
about $\om$-powers and in section 6 
 arithmetical properties of $\om$-CFL. Section 7 deals with context free 
languages of infinite trees.

\section{$\om$-regular and  $\om$-context free  \la s}

We assume the reader to be familiar with the theory of formal \la s and 
of \orl s, see for example \cite{hu69} ,\cite{tho}.
We first recall some of the definitions and results concerning $\om$-regular 
and $\om$-context free  \la s and omega pushdown automata as presented in \cite{tho} 
\cite{cg} , 
\cite{cgdet}.
\nl
When $\Si$ is a finite alphabet, a finite string (word) over $\Si$ is any sequence $x=x_1\ldots x_k$ , where $x_i\in\Sigma$ 
for $i=1, \ldots ,k$ ,and  $k$ is an integer $\geq 1$. The length
 of $x$ is $k$, denoted by $|x|$ .
\nl we write $x(i)=x_i$  and $x[i]=x(1) \ldots x(i)$ for $i\leq k$.
\nl
 If  $|x|=0$ , $x$ is the empty word denoted by $\lambda$.
\nl $\Sis$  is the set of finite words over $\Sigma$.
\nl  The first infinite ordinal is $\om$.
\nl An $\om$-word over $\Si$ is an $\om$ -sequence $a_1 \ldots a_n \ldots $, where 
$a_i \in\Sigma , \fa i\geq 1$.
\nl When $\sigma$ is an $\om$-word over $\Si$, we write
 $\sigma =\sigma(1)\sigma(2) \ldots \sigma(n) \ldots  $
\nl  $\sigma[n]=\sigma(1)\sigma(2) \ldots \sigma(n)$ is the finite word of length n, 
prefix of $\sigma$.
\nl The set of $\om$-words over  the alphabet $\Si$ is denoted by $\Si^\om$.
\nl An  $\om$-language over an alphabet $\Sigma$ is a subset of  $\Si^\om$.

 The usual concatenation product of two finite words $u$ and $v$ is 
denoted $u.v$ (and sometimes just $uv$). This product is extended to the product of a 
finite word $u$ and an $\om$-word $v$: the infinite word $u.v$ is then the $\om$-word such that:
\nl $(u.v)(k)=u(k)$  if $k\leq |u|$ , and 
\nl $(u.v)(k)=v(k-|u|)$  if $k>|u|$.

 For $V\subseteq \Sis$, $V^\om = \{ \sigma =u_1  \ldots u_n \ldots 
 \in \Si^\om \mid  u_i\in V, \fa i\geq 1 \}$
is the $\om$-power of $V$.
\nl For $V\subseteq \Sis$, the complement of $V$ (in $\Sis$) is $\Sis - V$ denoted $V^-$.
\nl  For a subset $A\subseteq \Si^\om$, the complement of $A$ is 
$\Si^\om - A$ denoted $A^-$.

    The prefix relation is denoted $\sqsubseteq$: the finite word $u$ is a prefix of the finite 
word $v$
(denoted $u\sqsubseteq v$) if and only if there exists a (finite) word $w$ such that $v=u.w$.
\nl This definition is extended to finite words which are prefixes of $\om$-words:
\nl the finite word $u$ is a prefix of the $\om$-word $v$ (denoted $u\sqsubseteq v$) 
iff there exists an $\om$-word $w$ such that $v=u.w$.

\begin{Deff} A finite state machine (FSM) is a quadruple $M=(K,\Si,\delta, q_0)$, where $K$ 
is a finite set of states, $\Sigma$ is a finite input alphabet, $q_0 \in K$ is the initial state
and $\delta$ is a mapping from $K \times   \Si$ into $2^K$ . A FSM is called deterministic
 (DFSM) iff :
$\delta : K \times  \Si \ra K$.
\nl 
A B\"uchi automaton (BA) is a 5-tuple $M=(K,\Si,\delta, q_0, F)$ where
  $M'=(K,\Si,\delta, q_0)$
is a finite state machine and $F\subseteq K$ is the set of final states.
\nl
A Muller automaton (MA) is a 5-tuple $M=(K,\Si,\delta, q_0, F)$ where
$M'=(K,\Si,\delta, q_0)$ is a FSM and $F\subseteq 2^K$ is the collection of 
designated state sets.
\nl
A B\"uchi or Muller automaton is said \de if the associated FSM is deterministic.
\nl
Let $\sigma =a_1a_2 \ldots a_n \ldots $ be an  $\om$-word over $\Si$.
 \nl 
A sequence of states $r=q_1q_2 \ldots q_n \ldots $  is called 
an (infinite) run of $M=(K,\Si,\delta, q_0)$ on $\sigma$, 
starting in state $p$, iff:
1) $q_1=p$  and 2) for each $i\geq 1$, $q_{i+1} \in \delta( q_i,a_i)$.
\nl
In case a run $r$ of $M$ on $\sigma$ starts in state $q_0$, we call it simply "a run of $M$ 
on $\sigma$ " .
\nl
For every (infinite) run $r=q_1q_2 \ldots q_n \ldots $ of $M$, $In(r)$ is the set of
states in $K$ entered by $M$ infinitely many times during run $r$:
\nl
$In(r)= \{ q\in K  \mid \{i\geq 1  \mid q_i=q\} $   is infinite $\}   $.
\nl For $M=(K,\Si,\delta, q_0, F)$ a BA ,
the \ol~ accepted by $M$ is
$L(M)= \{  \sigma\in\Si^\om \mid $ there exists a  run r
 of M on $\sigma$ such that $In(r) \cap F \neq\emptyset \}$.
\nl For $M=(K,\Si,\delta, q_0, F)$ a MA, the  \ol~ accepted by $M$ is 
$L(M)= \{  \sigma\in\Si^\om \mid $ there exists a  run r
 of M on $\sigma$ such that $In(r) \in F \}$.
\end{Deff}

\noi The classical result of R. Mc Naughton \cite{rmc}  established that the expressive
 power of \de MA (DMA) is equal to the expressive power of non \de MA
(NDMA) which is also equal to the expressive power of non \de BA (NDBA) .
\nl There is  also a characterization of  \la s accepted by MA by means 
of the "$\om$-Kleene closure" of which we give now the definition:

\begin{Deff}
For any family L of  finitary \la s over the alphabet $\Si$, the $\om$-Kleene closure
of L, is : $$\om-KC(L) = \{ \cup_{i=1}^n U_i.V_i^\om  \mid  U_i, V_i \in L , \fa i\in [1, n] \}$$
\end{Deff}

\begin{The}
  For any \ol~ $L$, the following conditions are equivalent:
\begin{enumerate}
\ite   $L$ belongs to $\om-KC(REG)$ , where $REG$ is the class of (finitary)
 regular languages.
\ite   There exists a DMA  that accepts $L$.
\ite   There exists a MA  that accepts $L$.
\ite     There exists a BA  that accepts $L$.
\end{enumerate}

\noi An \ol~ $L$ satisfying one of the conditions of the above Theorem is called 
 an \orl . The class of \orl s will
 be denoted by $REG_\om$.
\end{The}

\noi We now define  pushdown machines and the class of  $\om$-context free  \la s.

\begin{Deff}
A pushdown machine (PDM) is a 6-tuple $M=(K,\Si,\Ga, \delta, q_0, Z_0)$, where $K$ 
is a finite set of states, $\Sigma$ is a finite input alphabet, $\Gamma$ is a 
finite pushdown alphabet,
 $q_0\in K$ is the initial state, $Z_0 \in\Ga$ is the start symbol, 
and $\delta$ is a mapping from $K \times (\Si\cup\{\lambda\} )\times \Ga $ to finite subsets of
$K\times \Gas$ . 
\nl
If  $\gamma\in\Ga^{+}$ describes the pushdown store content, 
the leftmost symbol will be assumed to be on " top" of the store.
A configuration of a PDM is a pair $(q, \gamma)$ where $q\in K$ and  $\gamma\in\Gas$.\nl
For $a\in \Si\cup\{\lambda\}$, $\beta,\gamma \in\Ga^{\star}$
and $Z\in\Ga$, if $(p,\beta)$ is in $\delta(q,a,Z)$, then we write
$a: (q,Z\gamma)\mapsto_M (p,\beta\gamma)$.\nl
$\mapsto_M^\star$ is the transitive and reflexive closure of $\mapsto_M$.
(The subscript $M$ will be omitted whenever the meaning remains clear).
\nl
Let $\sigma =a_1a_2 \ldots a_n \ldots $ be an  $\om$-word over $\Si$. 
an infinite sequence of configurations $r=(q_i,\gamma_i)_{i\geq1}$ is called 
a complete run of $M$ on $\sigma$, starting in configuration $(p,\gamma)$, iff:
\begin{enumerate}
\ite $(q_1,\gamma_1)=(p,\gamma)$

\ite  for each $i\geq 1$, there exists $b_i\in\Si\cup\{\lambda\}$ 
satisfying $b_i: (q_i,\gamma_i)\mapsto_M(q_{i+1},\gamma_{i+1} )$
such that $a_1a_2 \ldots a_n  \ldots  =b_1b_2 \ldots b_n \ldots $
\end{enumerate}
\noi
As for FSM, for every such run, $In(r)$ is the set of all states entered infinitely
 often during run $r$.
\nl
A complete run $r$ of $M$ on $\sigma$ , starting in configuration $(q_0,Z_0)$,
 will be simply called " a run of $M$ on $\sigma$ ".
\end{Deff}

\begin{Deff} A B\"uchi pushdown automaton (BPDA) is a 7-tuple
 $M=(K,\Si,\Ga, \delta, q_0, Z_0, F)$ where $ M'=(K,\Si,\Ga, \delta, q_0, Z_0)$
is a PDM and $F\subseteq K$ is the set of final states.
\nl
The \ol~ accepted by $M$ is 
$L(M)= \{  \sigma\in\Si^\om \mid$ there exists a complete run r
 of M on $\sigma$ such that $In(r) \cap F \neq\emptyset \}$.
\end{Deff}

\begin{Deff} A Muller pushdown automaton (MPDA) is a 7-tuple
 $M=(K,\Si,\Gamma, \delta, q_0, Z_0, F)$ where $ M'=(K,\Si,\Gamma, \delta, q_0, Z_0)$
is a PDM and $F\subseteq 2^K$ is the collection of designated state sets.
\nl
The \ol~ accepted by $M$ is 
$L(M)= \{  \sigma\in\Si^\om \mid$ there exists a complete run r
 of M on $\sigma$ such that $In(r) \in F \}$.
\end{Deff}

\begin{Rem} We consider here two acceptance conditions for $\om$-words, 
the B\"uchi  and the Muller acceptance conditions, respectively denoted 2-acceptance 
and 3-acceptance in \cite{la} and in \cite{cgdet} and $(inf, \sqcap)$ and $(inf, =)$ 
in \cite{sta}.
\end{Rem}

\noi Cohen and  Gold and independently Linna established a characterization 
Theorem for $\om$-CFL:

\begin{The}\label{theokccf}
Let $CFL$ be the class of context free (finitary) languages. Then for 
any  \ol~ $L$ the following
three conditions are equivalent:
\begin{enumerate}
\ite $L\in \om -KC(CFL)$.
\ite There exists a $BPDA$ that accepts $L$.
\ite There exists a $MPDA$ that accepts $L$.
\end{enumerate}
\end{The}

\noi
In \cite{cg}  are also studied  \ol s generated by $\om$-context free grammars  
and it is shown that each of the conditions 1), 2), and 3) of the above Theorem is 
also equivalent to: 4) $L$ is generated by a context free grammar $G$ by leftmost derivations.
These grammars are also studied in \cite{ni77} \cite{ni78}.
\nl
Then we can let the following definition:

\begin{Deff}
An \ol~ is an $\om$-context free language ($\om$-CFL) (or context free \ol~) iff it satisfies 
one of the conditions of the above Theorem.
\end{Deff}

\section{Topology}

\noi We assume the reader to be familiar with basic notions of topology which
may be found in \cite{lt} \cite{pp} \cite{ku} \cite{mos} \cite{kec}.

Topology is an important tool for the study of \ol s, and leads 
to characterization of several classes of \ol s.
\nl For a finite alphabet $X$, we consider $X^\om$ 
as a topological space with the Cantor topology.
 The open sets of $X^\om$ are the sets in the form $W.X^\om$, where $W\subseteq X^\star$.
A set $L\subseteq X^\om$ is a closed set iff its complement $X^\om - L$ is an open set.
The class of open sets of $X^\om$ will be denoted by ${\bf G}$ or by ${\bf \Si^0_1 }$. 
The class of closed sets will be denoted by ${\bf F}$ or by ${\bf \Pi^0_1 }$. 
\noi Define now the next classes of the Borel Hierarchy:

\begin{Deff}
The classes ${\bf \Si_n^0}$ and ${\bf \Pi_n^0 }$ of the Borel Hierarchy
 on the topological space $X^\om$  are defined as follows:
\nl ${\bf \Si^0_1 }$ is the class of open sets of $X^\om$.
\nl ${\bf \Pi^0_1 }$ is the class of closed sets of $X^\om$.
\nl ${\bf \Pi^0_2 }$  or ${\bf G_\delta }$ is the class of countable intersections of 
 open sets of $X^\om$.
\nl  ${\bf \Si^0_2 }$  or ${\bf F_\sigma }$ is the class of countable unions  of 
closed sets of $X^\om$.
\nl And for any integer $n\geq 1$:
\nl ${\bf \Si^0_{n+1} }$   is the class of countable unions 
of ${\bf \Pi^0_n }$-subsets of  $X^\om$.
\nl ${\bf \Pi^0_{n+1} }$ is the class of countable intersections of 
${\bf \Si^0_n}$-subsets of $X^\om$.
\nl The Borel Hierarchy is also defined for transfinite levels.
The classes ${\bf \Si^0_\alpha }$
 and ${\bf \Pi^0_\alpha }$, for a countable ordinal $\alpha$, are defined in the
 following way:
\nl ${\bf \Si^0_\alpha }$ is the class of countable unions of subsets of $X^\om$ in 
$\cup_{\gamma <\alpha}{\bf \Pi^0_\gamma }$.
 \nl ${\bf \Pi^0_\alpha }$ is the class of countable intersections of subsets of $X^\om$ in 
$\cup_{\gamma <\alpha}{\bf \Si^0_\gamma }$.
\end{Deff}

\noi Recall some basic results about these classes, \cite{mos}:

\begin{Pro}
\noi  
\begin{enumerate}
\ite[(a)] ${\bf \Si^0_\alpha }\cup {\bf \Pi^0_\alpha } \subsetneq  
{\bf \Si^0_{\alpha +1}}\cap {\bf \Pi^0_{\alpha +1} }$, for each countable 
ordinal  $\alpha \geq 1$. 
\ite[(b)] $\cup_{\gamma <\alpha}{\bf \Si^0_\gamma }= \cup_{\gamma <\alpha}{\bf \Pi^0_\gamma }
\subsetneq {\bf \Si^0_\alpha }\cap {\bf \Pi^0_\alpha }$, for each countable limit ordinal 
$\alpha$. 
\ite[(c)] A set $W\subseteq X^\om$ is in the class ${\bf \Si^0_\alpha }$ iff its 
complement is in the class ${\bf \Pi^0_\alpha }$. 
\ite[(d)] ${\bf \Si^0_\alpha } - {\bf \Pi^0_\alpha } \neq \emptyset $ and 
${\bf \Pi^0_\alpha } - {\bf \Si^0_\alpha } \neq \emptyset $ hold 
 for every countable  ordinal $\alpha\geq 1$. 
\end{enumerate}
\end{Pro}

\noi  We shall say that a subset of $X^\om$ is a Borel set of rank $\alpha$, for 
a countable ordinal $\alpha$,  iff 
it is in ${\bf \Si^0_{\alpha}}\cup {\bf \Pi^0_{\alpha}}$ but not in 
$\bigcup_{\gamma <\alpha}({\bf \Si^0_\gamma }\cup {\bf \Pi^0_\gamma})$. 

    Furthermore, when $X$ is a finite set, there are some subsets of $X^\om$ which 
are not Borel sets.
Indeed there exists another hierarchy beyond the Borel hierarchy, which is called the 
projective hierarchy and which is obtained from  the Borel hierarchy by 
successive applications of operations of projection and complementation.
More precisely, a subset $A$ of  $X^\om$ is in the class ${\bf \Si^1_1}$ of {\bf analytic} sets
iff there exists another finite set $Y$ and a Borel subset $B$  of  $(X\times Y)^\om$ 
such that $ x \in A \lra \exists y \in Y^\om $ such that $(x, y) \in B$.
\nl Where $(x, y)$ is the infinite word over the alphabet $X\times Y$ such that
$(x, y)(i)=(x(i),y(i))$ for each  integer $i\geq 0$.
\nl Now a subset of $X^\om$ is in the class ${\bf \Pi^1_1}$ of {\bf coanalytic} sets
iff its complement in $X^\om$ is an analytic set.
\nl The next classes are defined in the same manner, ${\bf \Si^1_{n+1}}$-sets of  
$X^\om$ are projections of ${\bf \Pi^1_n}$-sets and  ${\bf \Pi^1_{n+1}}$-sets 
are the complements of ${\bf \Si^1_{n+1}}$-sets.

    Recall also the notion of completeness with regard to reduction by continuous functions. 
\nl A set $F\subseteq X^\om$ is a ${\bf \Si^0_\alpha}$
 (respectively ${\bf \Pi^0_\alpha}$)-complete set iff for any set $E\subseteq Y^\om$
($Y$ a finite alphabet): 
\nl $E\in {\bf \Si^0_\alpha}$ (respectively $E\in {\bf \Pi^0_\alpha}$) 
iff there exists a continuous 
function $f$ from $Y^\om$ into $X^\om$ such that $E = f^{-1}(F)$. 
\nl A similar notion exists for classes of the projective hierarchy: in particular 
a set $F\subseteq X^\om$ is a ${\bf \Si^1_1}$
 (respectively ${\bf \Pi^1_1 }$)-complete set iff for any set $E\subseteq Y^\om$
($Y$ a finite alphabet): 
\nl $E\in {\bf \Si^1_1}$ (respectively $E\in {\bf \Pi^1_1}$) iff there exists a continuous 
function $f$ from $Y^\om$ into $X^\om$ such that $E = f^{-1}(F)$. 

    A ${\bf \Si^0_\alpha}$
 (respectively ${\bf \Pi^0_\alpha}$, ${\bf \Si^1_1}$)-complete set is a ${\bf \Si^0_\alpha}$
 (respectively ${\bf \Pi^0_\alpha}$, ${\bf \Si^1_1}$)- set which is in some 
sense a set of the highest 
topological complexity among the ${\bf \Si^0_\alpha}$
 (respectively ${\bf \Pi^0_\alpha}$, ${\bf \Si^1_1}$)- sets.

\section{topological properties of $\om$-CFL}

\noi Recall first previous results.
  $\om$-CFL exhaust the finite ranks of the Borel hierarchy.

\begin{The}[\cite{fina}]
For each  integer $n\geq 1$, there exist some 
${\bf \Si_n^0}$-complete $\om$-CFL and some 
${\bf \Pi_n^0}$-complete $\om$-CFL.
\end{The}

\noi Cohen and Gold proved that one cannot decide whether an $\om$-CFL is in the class 
${\bf F, G }$ or ${\bf G_\delta}$. We have extended in \cite{fina} 
this result to all classes ${\bf \Si_n^0}$ and
${\bf \Pi_n^0}$, for n an integer $\geq 1$. (We say that an $\om$-CFL $A$  
is effectively given when a MPDA accepting 
$A$ is given).

\begin{The}[\cite{fina}]\label{indbor}
Let n be an integer $\geq 1$. Then it is undecidable whether an effectively given $\om$-CFL
is in the class ${\bf \Si_n^0}$ ( repectively ${\bf \Pi_n^0}$).
\end{The}

    When considering $\om$-CFL, natural questions now arise: are all $\om$-CFL
Borel sets of finite rank, Borel sets, analytic sets....?
First recall the following:

\begin{The}[\cite{sta}]\label{cfana}
Every $\om$-CFL over a finite alphabet $X$ is an analytic subset of $X^\om$.
\end{The}

\proo we just sketch the proof.
\nl Every   $\om$-CFL $A \subseteq \Si^\om$ is the projection 
of a \de $\om$-CFL onto $\Si^\om$ but 
 \de $\om$-CFL are Borel sets of rank at most 3, and it is well
known that such a projection of a Borel set is an analytic subset of $\Si^\om$. 
Remark that in fact each $\om$-CFL is the projection of an $\om$-CFL which is accepted by a 
\de B\"uchi pushdown automaton and therefore  which is a ${\bf \Pi^0_2}$-set. \ep

\begin{Rem}
This above theorem is in fact true for \ol s accepted by Turing machines which are 
much more powerful accepting devices than pushdown automata \cite{sta}.
\end{Rem}

    The following question now  arises: are there $\om$-CFL  which are analytic 
but not Borel sets?

\begin{The}\label{cfnotbor}
There exist $\om$-CFL which are ${\bf \Si^1_1}$-complete hence non Borel sets.
\end{The}

\proo  We shall use here results about languages of infinite binary trees whose nodes
are labelled in a finite alphabet $\Si$.
\nl A node of an infinite binary tree is represented by a finite  word over 
the alphabet $\{l, r\}$ where $r$ means "right" and $l$ means "left". Then an 
infinite binary tree whose nodes are labelled  in $\Si$ is identified with a function
$t: \{l, r\}^\star \ra \Si$. The set of  infinite binary trees labelled in $\Si$ will be 
denoted $T_\Si^\om$.

     There is a natural topology on this set $T_\Si^\om$ \cite{mos} \cite{lt}\cite{sim}. 
It is defined 
by the following distance. Let $t$ and $s$ be two distinct infinite trees in $T_\Si^\om$. 
Then the distance between $t$ and $s$ is $\frac{1}{2^n}$ where $n$ is the smallest integer 
such that $t(x)\neq s(x)$ for some word $x\in \{l, r\}^\star$ of length $n$.
\nl The open sets are then in the form $T_0.T_\Si^\om$ where $T_0$ is a set of finite labelled
trees. $T_0.T_\Si^\om$ is the set of infinite binary trees 
which extend some finite labelled binary tree $t_0\in T_0$, $t_0$ is here a sort of prefix, 
an "initial subtree"
of a tree in $t_0.T_\Si^\om$.

    The Borel hierarchy and the projective hierarchy on $T_\Si^\om$ are defined from open 
sets in the same manner as in the case of the topological space $\Si^\om$.

    Let $t$ be a tree. A branch $B$ of $t$ is a subset of the set of nodes of $t$ which 
is linearly ordered by the tree partial order $\sqsubseteq$ and which 
is closed under prefix relation, 
i.e. if  $x$ and $y$ are nodes of $t$ such that $y\in B$ and $x \sqsubseteq y$ then $x\in B$.
\nl A branch $B$ of a tree is said to be maximal iff there is not any other branch of $t$ 
which strictly contains $B$.

    Let $t$ be an infinite binary tree in $T_\Si^\om$. If $B$ is a maximal branch of $t$,
then this branch is infinite. Let $(u_i)_{i\geq 0}$ be the enumeration of the nodes in $B$
which is strictly increasing for the prefix order. 
\nl  The infinite sequence of labels of the nodes of  such a maximal 
branch $B$, i.e. $t(u_0)t(u_1) \ldots t(u_n) \ldots $  is called a path. It is an $\om$-word 
over the alphabet $\Si$.

    Let then $L\subseteq \Si^\om$ be an \ol~ over $\Si$. Then  we denote $Path(L)$  the set of 
infinite trees $t$ in $T_\Si^\om$ such that $t$ has (at least) one path in $L$.

    It is well known that if $L\subseteq \Si^\om$ is an \ol~ over $\Si$ which is a 
${\bf \Pi^0_2 }$-complete subset of $\Si^\om$ (or a set of higher complexity in the Borel 
hierarchy) then the set $Path(L)$  is a ${\bf \Si^1_1 }$-complete subset of $T_\Si^\om$. 
Hence $Path(L)$  is not a Borel set, \cite{niw85} \cite{sim} \cite{simcras}.

    Whenever $B$  is an $\om$-CFL we shall find another 
$\om$-CFL  $C$ and a continuous function 

$$h: T_\Si^\om   \ra  (\Si\cup\{A\})^\om $$

\noi such that  $Path(B) = h^{-1} ( C )$.  
For that we will code trees labelled in $\Si$ by  words over $\Si \cup\{A\}=\Si_A$, where 
$A$ is  supposed to be a new letter not in $\Si$. 

    Consider now the set $\{l, r\}^\star$ of nodes of binary infinite trees.
For each integer $n\geq 0$, call $C_n$ the set of words of length $n$ of $\{l, r\}$. 
Then $C_0=\{\lambda\}$, $C_1=\{l, r\}$, $C_2=\{ll, lr, rl, rr\}$ and so on.
$C_n$ is the set of nodes which appear in the $(n+1)$th level of an infinite binary tree.
The number of nodes of $C_n$ is $card(C_n)=2^n$.  We consider now  
the lexicographic order on $C_n$ (assuming that $l$ is before $r$ for this order).
Then, in the enumeration of the nodes with regard to this order, the  nodes of $C_1$ will 
be: $l, r$; the nodes of $C_3$ will be: $ lll, llr, lrl, lrr, rll, rlr, rrl, rrr$.
\nl Let $u^n_1, \ldots , u^n_j,  \ldots , u^n_{2^n}$ be such an enumeration of $C_n$ in 
the lexicographic order and let $v^n_1, \ldots , v^n_j, \ldots , v^n_{2^n}$ 
be the enumeration of the 
elements
of $C_n$ in the reverse order. Then for all integers $n\geq 0$ and $i$, $1\leq i\leq 2^n$, 
it holds that $v_i^n=u^n_{2^n+1-i}$.

    We define now the code of a tree $t$ in $ T_\Si^\om$.
Let $A$ be a  letter not in $\Si$. We construct an $\om$-word over the alphabet 
$(\Si\cup\{A\})$ which will code the tree $t$.  We enumerate all the labels of the 
nodes of a tree in the following manner:
firstly the label of the node of $C_0$ which is $t(u_1^0)$,
\nl followed by an $A$, followed by the labels of nodes of $C_1$ in the lexicographic order, 
i.e. $t(u_1^1)t(u_2^1)$, followed by an $A$, followed by the labels of the nodes 
of $C_2$ in the reverse lexicographic order, followed by an $A$, 
 followed by the labels of nodes of $C_3$ in the lexicographic order, and so on \ldots
\nl For each integer $n\geq 0$, the labels of the nodes of $C_n$ are enumerated before 
those of $C_{n+1}$ and these two 
sets of labels are separated by an $A$. Moreover the labels of the nodes of $C_{2n+1}$, for 
$n\geq 0$, are enumerated  in the lexicographic order (for the nodes) and 
the labels of the nodes of $C_{2n}$,  for 
$n\geq 0$, are enumerated  in the reverse lexicographic order (for the nodes).

    Then for each tree $t$ in $ T_\Si^\om$, we obtain an $\om$-word of 
$\Si\cup\{A\}$ which will be denoted $h(t)$. With the preceding notations it holds that:

$$h(t)=t(u_1^0)At(u_1^1)t(u_2^1)At(v_1^2)t(v_2^2)t(v_3^2)t(v_4^2)A
t(u_1^3)t(u_2^3)t(u_3^3)t(u_4^3)t(u_5^3)t(u_6^3)t(u_7^3)t(u_8^3)A \ldots $$

    Let then $h$ be the mapping from $T_\Si^\om$ into $(\Si \cup \{A\})^\om$ 
such that for every labelled binary infinite tree $t$ of $T_\Si^\om$, 
$h(t)$ is the code of the tree as defined above.
It is easy to see, from the definition of $h$ and of the  order of the enumeration 
of labels of nodes, that $h$ is a continuous function from $T_\Si^\om$ into 
$(\Si \cup \{A\})^\om$.

    Assume now that $B$ is an $\om$-CFL accepted by a 
 B\"uchi pushdown automaton 
 $M=(K,\Si,\Ga, \delta, q_0, Z_0, F)$ where $ M'=(K,\Si,\Ga, \delta, q_0, Z_0)$
is a pushdown machine and $F\subseteq K$ is the set of final states.

    Now we are looking for another $\om$-CFL $C$ such that 
for every  tree $t \in T_\Si^\om$,  $h(t) \in C$ if and only if  $t$ has
a path in  $B$. Then  we shall have $Path(B) = h^{-1} ( C )$.

We shall give a {\bf first description} of such an  $\om$-CFL $C$ by 
constructing  from $M$ another B\"uchi pushdown automaton $\bar{M}$
 which accepts $C$.
\nl The reader can also skip this description and read a {\bf second description }
of the $\om$-CFL $C$ which will be given below. 

Describe first informally  the behaviour of the new machine $\bar{M}$.
When ${\bar M}$ reads a word in the form $h(t)$, then using the non determinism 
it guesses a maximal branch of the tree $t$
and simulates on this branch the B\"uchi pushdown automaton $M$. Finally 
the acceptation of $h(t)$ by ${\bar M}$ is related to the acceptation of the 
$\om$-word formed  by the labels of this branch by $M$. 

    More formally 
$\bar{M}=(\bar{K},\bar{\Si},\bar{\Ga}, \bar{\delta}, \bar{q_0}, \bar{Z_0}, \bar{F})$,
 where 
 $$\bar{K}=K \cup \{q^1 \mid q\in K\} \cup \{q^2 \mid q\in K\} \cup \{q^3 \mid q\in K\}
\cup \{q^4 \mid q\in K\} \cup \{q^5 \mid q\in K\} \cup\{q_r\} $$ 
$$\bar{\Si}=\Si \cup \{A\}$$
$$\bar{\Ga}=\Ga \cup \{E\}$$
\noi where $E$ is a new letter not in $\Ga$,
$$\bar{q_0}=q_0$$
$$\bar{Z_0}=Z_0$$
$$\bar{F}=F\cup\{q^5 \mid q\in F\}$$
\noi and the transition relation $\bar{\delta}$ is defined by the following cases
which will be explained below:

\begin{enumerate}
\ite[(a)] $(q, \nu) \in \bar{\delta}(q_0, a, Z_0)$ iff 
$(q, \nu) \in \delta(q_0, a, Z_0)$, for each $a\in \Si$ and $\nu \in \Ga^\star$.
\ite[(b)]  $\bar{\delta}(q_0, A, Z_0)=(q_r, Z_0)$. 
\ite[(c)]  $\bar{\delta}(q, a, Z)=(q^1, EZ)$, for each $a\in \Si$, 
$Z\in \Ga\cup\{E\}$ and $q\in K$.
\ite[(d)]  $\bar{\delta}(q^1, a, E)=(q^1, EE)$, for each $a\in \Si$, 
 and $q\in K$.

\ite[(e)]  $\bar{\delta}(q^1, A, Z)=(q^2, Z)$, for each $Z\in \Ga\cup\{E\}$ and $q\in K$.
\ite[(f)]  $\bar{\delta}(q, A, Z)=(q^2, Z)$, for each $Z\in \Ga\cup\{E\}$ and $q\in K$.

\ite[(g)]  $\bar{\delta}(q^2, a, E)=(q^3, E)$, for each $a\in \Si$, and $q\in K$.
\ite[(h)]  $\bar{\delta}(q^3, a, E)=(q^2, \lambda)$, for each $a\in \Si$, and $q\in K$.
\ite[(i)]  $\bar{\delta}(q^2, A, E)=(q_r, E)$,  for each $q\in K$.
\ite[(j)]  $\bar{\delta}(q^3, A, E)=(q_r, E)$,  for each $q\in K$.
\ite[(k)]  $\bar{\delta}(q_r, a, Z)=(q_r, Z)$,  for each $a\in (\Si\cup\{A\})$ and 
$Z\in \Ga\cup\{E\}$.
\ite[(l)]  $\bar{\delta}(q^2, a, Z) \ni (q', \nu)$ iff   
           $\delta(q, a, Z)\ni (q', \nu)$, for each $a\in \Si$, $q, q'\in K$, $Z\in \Ga$, and 
           $\nu \in \Ga^\star$.
\ite[(m)]  $\bar{\delta}(q^2, \lambda, Z) \ni (q'^5, \nu)$ iff   
           $\delta(q, \lambda, Z)\ni (q', \nu)$, for each  $q, q'\in K$, $Z\in \Ga$, and 
           $\nu \in \Ga^\star$.
\ite[(n)]  $\bar{\delta}(q^5, \lambda, Z) \ni (q'^5, \nu)$ iff   
           $\delta(q, \lambda, Z)\ni (q', \nu)$, for each  $q, q'\in K$, $Z\in \Ga$, and 
           $\nu \in \Ga^\star$.
\ite[(o)]  $\bar{\delta}(q^5, a, Z) \ni (q', \nu)$ iff   
           $\delta(q, a, Z)\ni (q', \nu)$, for each $a\in \Si$, $q, q'\in K$, $Z\in \Ga$, and 
           $\nu \in \Ga^\star$.
\ite[(p)]  $\bar{\delta}(q^5, a, Z) \ni (q^4, Z)$, for each $a\in \Si$, 
$Z\in \Ga$ and $q\in K$.

\ite[(q)]  $\bar{\delta}(q^2, a, Z) \ni (q^4, Z)$, for each $a\in \Si$, $Z\in \Ga$ and $q\in K$.
\ite[(r)]  $\bar{\delta}(q^4, a, Z) \ni (q', \nu)$ iff  $\delta(q, a, Z)\ni (q', \nu)$, 
for each $a\in \Si$, $q, q'\in K$, $Z\in \Ga$, and 
           $\nu \in \Ga^\star$.

\ite[(s)]  $\bar{\delta}(q^4, A, Z)= (q_r, Z)$,  for each $q\in K$, $Z\in \Ga$.
\end{enumerate}

    We describe now more precisely the behaviour of $\bar{M}$.
\nl  To the set $K$ of states of $M$, we add sets of states $K^{i}=\{q^{i} \mid q\in K\}$
for each integer $i\in [1, 5]$, and a state $q_r$ which will be a rejecting state.

    We firstly consider only the reading by $\bar{M}$ of words in the form $h(t)$ where 
$t\in T_\Si^\om$. 
When $\bar{M}$  simulates $M$ on the branch it guesses, it enters in a state of $K$,
as indicated by  $(a), (l), (o), (r)$,
or of $K^5$ if it uses a $\lambda$-transition, i.e. if it does not read any letter 
during this transition, as indicated by $(m)-(n)$. 

 When $\bar{M}$  reads the labels of the nodes of $t$, it reads successively  the labels 
of nodes of $C_0, C_1, C_2, \ldots , C_i, \ldots $

 Let $B$ be the branch which is  guessed by $\bar{M}$ during a reading. 

    After the use of one transition rule of $(a), (l), (o)$ or $(r)$, reading the label 
of a node $u$ of $B$ in $C_n$, $n\geq 0$, 
$\bar{M}$ enters in a  state $q^1$, keeping the memory of $q$, and then continues the 
reading of the (labels of) nodes of $C_n$, pushing an $E$ on the top of the stack for every 
letter of $\Si$ it reads (transition rules (c), (d)) until it reads an $A$.
Then it enters in state $q^2$, keeping again the memory of $q$, (transition rules (e), (f)) 
and reading the labels of nodes of $C_{n+1}$, it begins to pop an $E$ from the top of the stack
for two letters of $\Si$ it reads, as indicated by transition rules 
$(g)$, $(h)$ (here are used the two sets of states $K^2$ and $K^3$). 
Thus when the letter at the top of the stack is again a 
letter of $\Ga$ (and not an $E$) the machine $\bar{M}$ reads the label of one successor of 
the node $u$ (this is due to the fact that the tree is binary and to the order of the 
enumeration of the nodes we have chosen in the definition of $h(t)$). 
It may choose to simulate $M$ on this label, as indicated by the transition 
rules $(l),(m),(n),(o)$,  (perhaps after some $\lambda$-transitions). Otherwise it may choose 
to wait the next label, entering in state $q^4$, as indicated by the transition rules 
$(p), (q)$, and then simulates $M$ as indicated by the transition rule $(r)$.

    Some other transition rules, $(b), (i), (j), (k), (s)$, lead to the rejecting state 
$q_r$ in which $\bar{M}$ remains for the rest of the reading. But in fact these 
transition rules are never used for the reading of $\om$-words in the form $h(t)$ where
$t\in T_\Si^\om$. 

    Now we can see that when $\bar{M}$ simulates $M$ on the branch $B$, if $M$ 
 enters in a state $q\in K$, then $\bar{M}$  enters in the  state $q$ or in the state $q^5$ 
(when a $\lambda$-transition is used).
Thus the choice of the set of accepting states $\bar{F}=F\cup\{q^5 \mid q\in F\}$
implies the property:
 for a tree $t \in T_\Si^\om$,  $h(t) \in C$ if and only if  $t$ has
a path in  $B$.

We are going now to give a {\bf second  description} of the  $\om$-CFL $C$.

The \ol~ $C$ which we have constructed from the 
\ol~ $B$ can easily be described by means of substitution of context free languages.  
\nl Let first $D$ be the following finitary language over the alphabet $(\Si \cup \{A\})$:
$$D=\{ u.A.v ~\mid ~ u, v \in \Sis ~and~ ( |v|=2|u|)~~ or ~~( |v|=2|u|+1)~ \}$$
\noi It is easy to see that $D$ is a context free language. 

Now an $\om$-word $\sigma\in C$ may be considered as an  $\om$-word
$\sigma' \in B$ to which we  add, 
  between two consecutive letters $\sigma'(n)$ and $\sigma'(n+1)$
 of $\sigma'$, a finite word  $v_{n}$  belonging to the context free finitary 
language $D$.

 Recall now the definition of substitution in languages:
 A substitution $f$ is defined by a mapping 
$\Si\ra P(\Ga^\star)$, where $\Si =\{a_1, \ldots ,a_n\}$  and $\Ga$ are two finite alphabets, 
$f: a_i \ra L_i$ where $\fa i\in [1;n]$, $L_i$ is a finitary language over the alphabet $\Ga$.
\nl Now this mapping is extended in the usual manner to finite words:
$$f(x(1) \ldots x(n))= \{u_1 \ldots u_n \quad \mid \quad u_i\in f(x(i)) , ~\fa i\in [1;n]\}$$ 
\noi where $x(1)$, \ldots , $x(n)$ are letters in $\Si$, 
and to finitary languages $L\subseteq \Sis$: 
$$f(L)=\cup_{x\in L} f(x)$$ 
\noi The substitution $f$ is called $\lambda$-free if $\fa i\in [1;n]$  $L_i$ does not 
contain the empty word. In that case the mapping $f$ may be extended to $\om$-words:
$$f(x(1)\ldots x(n)\ldots )= \{u_1\ldots u_n \ldots \quad \mid \quad u_i\in f(x(i)), ~ 
\fa i \geq 1 \}$$ 

\noi  Let $\mathbb{C}$ be a family of languages, if $\fa i\in [1;n]$ the language 
$L_i$ belongs to  $\mathbb{C}$ 
the substitution $f$ is called a $\mathbb{C}$-substitution.

 Let then $g$ be the substitution $\Si\ra P((\Si \cup \{A\})^\star)$ defined by:
$a \ra a.D$ where $D$ is the context free language defined above. Then $g$ is a 
$\lambda$-free substitution and $g(B)=C$ holds. But the languages $a.D$ are context free 
and $CFL_\om$ is closed under $\lambda$-free context free substitution \cite{cg}. 
Then $B \in  CFL_\om$ implies that $C \in  CFL_\om$.

Hence if  $B$ is a Borel set which is  a ${\bf \Pi_2^0}$-complete subset 
of $\Si^\om$ (or a set of higher complexity in the Borel hierarchy), the language
$h^{-1}(C)=Path(B)$ is a ${\bf \Si^1_1}$-complete subset of $T_\Si^\om$.
Then the \ol~  $C$ is at least  ${\bf \Si^1_1}$-complete because $h$ 
is a continuous function (note that here $h$ is a continuous function: $T_\Si^\om \ra (\Si_A)^\om$ 
and the preceding definition of ${\bf \Si^1_1}$-complete set involves continuous reductions: 
 $X^\om \ra Y^\om$; but the two topological spaces $T_\Si^\om $ and $(\Si_A)^\om$ have 
good similar properties which enable to extend the previous definition  to this new case 
\cite{mos}\cite{kec}). 
And $C$  is in fact a ${\bf \Si^1_1}$-complete subset of $(\Si \cup \{A\})^\om$ because every 
 $\om$-CFL  is an analytic set by  Theorem \ref{cfana}.
\nl Then in that case $C$ is not a Borel set because a ${\bf \Si^1_1}$-complete 
set is not a Borel set \cite{ku}\cite{mos}.
\nl Indeed this gives infinitely many non Borel  $\om$-CFL , because there exist 
infinitely many  $\om$-CFL  of borel rank $>2$.   \ep 

 Remark that in the above proof, whenever $B$ is an \orl~accepted by a B\"uchi automaton 
$M$, the resulting machine $\bar{M}$ is just a one counter machine, i.e. a pushdown machine 
having a stack alphabet $\bar{\Ga}=\{Z_0, E\}$, where $Z_0$ is the bottom symbol 
which always remains at the bottom of the pushdown store and appears only there. Then 
at any moment of any computation the word in the pushdown store is in the form 
$E^nZ_0$ where $n$ is an integer $\geq 0$. Thus it holds that:

\begin{Cor}
There exist one counter \ol s  which are ${\bf \Si^1_1}$-complete hence non Borel sets.
\end{Cor}

\noi Now we can deduce from the preceding  proof  the following undecidability result:

\begin{The}\label{cfindbor} Let $\Si$ be an alphabet containing at least two letters.
It is undecidable, for an effectively given  $\om$-CFL $B$  to determine whether 
$B$ is a Borel subset of $\Si^\om$.
\end{The}

\proo  Remark first that $h(T_\Si^\om)$ is the set of $\om$-words in $(\Si_A)^\om $
which belong to 
$$\Si.A.\Si^2.A.\Si^4.A.\Si^8.A \ldots A.\Si^{2^n}.A\Si^{2^{n+1}} \ldots $$

\noi In other words this is the set of words in $(\Si_A)^\om $ which contain infinitely many 
occurrences of the letter $A$, and have $2^n$ letters of $\Si$ between the n$th$ and 
the (n+1)$th$ occurrences of the letter $A$.
We shall first state the following:

\begin{Lem}
Let $\Si$ be a finite  alphabet. Then $(\Si_A)^\om - h(T_\Si^\om)$ is an omega 
context free language.
\end{Lem}

\proo Let $$A_1=(A \cup \Si^2 \cup \Si.A.A \cup \Si.A.\Si.A \cup \Si.A.\Si^3). (\Si_A)^\om$$

\noi $A_1$ is the set of words in $(\Si_A)^\om$ which have not any word of 
$\Si.A.\Si^2.A$ as prefix. 
$A_1$ is clearly an \orl~ hence it is also an $\om$-CFL.

Let now $B_1$ be  the set of finite words over the alphabet $\Si_A$ which are in the form 
$A.u.A.v.A $ where $u, v \in \Si$ and $|v|<2|u|$.
\noi And let $B_2$ be  the set of finite words over the alphabet $\Si_A$ which are in the form 
$A.u.A.v $ where $u, v \in \Si$ and $|v|>2|u|$.

Then it is easy to see that $B_1$ and $B_2$ are context free finitary languages, 
thus the \ol~ 

$$A_2 = [(\Si_A)^\star.B_1.(\Si_A)^\om]  \cup [(\Si_A)^\star.B_2.(\Si_A)^\om ]$$

\noi is an  an omega context free language by Theorem \ref{theokccf}.
\nl But $(\Si_A)^\om - h(T_\Si^\om) = A_1 \cup A_2$ and the class of context free 
\ol s is closed under union \cite{cg} therefore 
$(\Si_A)^\om - h(T_\Si^\om)$ is an omega 
context free language.  \ep 

 We recall now a result established in  \cite{fina} in the course of the proof of 
the above Theorem
\ref{indbor}. 
We had seen that: 

\begin{Lem}\label{lemfina}
There exists a family of (effectively given)  context free \ol s  
 $(A_{X,Y}^\sim)^d$ 
over the alphabet $\{a, b, c, \tla , d\}$ such that $(A_{X,Y}^\sim)^d$ is either 
 $\{a, b, c, \tla , d\}^\om$ or an \ol~ which is a Borel set but neither a 
${\bf \Pi_2^0}$-subset nor a ${\bf \Si_2^0}$-subset of $\{a, b, c, \tla , d\}^\om$.
But one cannot decide which case holds.
\end{Lem}

\noi Consider now these languages. Denote $B(X,Y)=(A_{X,Y}^\sim)^d$  and  
$\Si=\{a, b, c, \tla , d\}$.  
\nl Then there are two cases.
\nl In the first case $B(X,Y)=\Si^\om$.
\nl In the second case $B(X,Y)$ is neither a ${\bf \Pi_2^0}$-subset 
nor a ${\bf \Si_2^0}$-subset of $\Si^\om$. 

    Return now to the previous proof. 
\nl  In the first case  $Path(B(X,Y))=Path(\Si^\om)= T_\Si^\om$.
\nl  In the second case $Path(B(X,Y))$ is a ${\bf \Si^1_1}$-complete subset of $T_\Si^\om$.

    Construct now from $B(X,Y)$ another omega 
context free language $C(X,Y)$ over the alphabet $\Si_A$ 
in the same manner as we have constructed 
$C$ from $B$ in the above proof.

    Let then $D(X,Y)=C(X,Y) \cup [(\Si_A)^\om - h(T_\Si^\om)]$. 
 $D(X,Y)$ is an $\om$-CFL because it is the union of two $\om$-CFL 
and the class of omega context free \la s is closed under union.

    Then  two  cases may happen.
\nl In the first case, $Path(B(X,Y))= T_\Si^\om$ hence $h(T_\Si^\om)\subseteq C(X,Y)$
and $D(X,Y)=(\Si_A)^\om$. Therefore $D(X,Y)$ is a closed and open subset of 
$(\Si_A)^\om$. 

    In the second case $h^{-1}(D(X,Y))=h^{-1}(C(X,Y))=Path(B(X,Y))$ holds by 
construction and then $D(X,Y)$ is a ${\bf \Si^1_1}$-complete subset of $(\Si_A)^\om$, 
for the same reason as $C(X,Y)$ is ${\bf \Si^1_1}$-complete .

    But one cannot decide which case holds hence one cannot decide 
whether the context free \ol~ is a Borel set. 

    To see that the result is also true for an alphabet containing two letters,
consider the morphism $g: \{a, b, c, \tla , d, A\}^\star \ra \{a, b\}^\star$ defined 
by: $a\ra bab$, $b\ra ba^2b$, $c\ra ba^3b$, $(\tla) \ra ba^4b$, $d\ra ba^5b$, $A\ra ba^6b$.
\nl This morphism is $\lambda$-free and may be extended to infinite words in an obvious manner,
giving a continuous function $\bar{g}: \{a, b, c, \tla , d, A\}^\om \ra \{a, b\}^\om$.

    Let then $ F(X,Y)=\bar{g}(D(X,Y))$.  
\nl $ F(X,Y)$ is an $\om$-CFL because $D(X,Y)$ is an $\om$-CFL and the class of context free 
\ol s is closed under $\lambda$-free morphism \cite{cg}.
\nl There are again two cases.
\nl In the first case, $D(X,Y)=(\Si_A)^\om$, hence $D(X,Y)$ is a compact set and, the 
image of a compact set by a continuous function being a compact set,  
$ F(X,Y)=\bar{g}(D(X,Y))$ is a compact subset of $\{a, b\}^\om$, therefore it is a 
closed subset of $\{a, b\}^\om$.

    In the second case, $D(X,Y)=\bar{g}^{-1}(F(X,Y))$ and $D(X,Y)$ is
a ${\bf \Si^1_1}$-complete subset of $T_\Si^\om$, thus $ F(X,Y)$ is also 
at least a ${\bf \Si^1_1}$-complete subset of $\{a, b\}^\om$, and in fact it is a 
${\bf \Si^1_1}$-complete subset because it is an analytic set as an $\om$-CFL.  \ep 

    Remark that we have also extended Theorem \ref{indbor} to all Borel classes:

\begin{The}\label{indbor2}
Let $\alpha$ be a countable ordinal $\geq 1$. 
Then it is undecidable to determine whether an effectively given $\om$-CFL
is in the class ${\bf \Si_{\alpha}^0}$ ( respectively ${\bf \Pi_{\alpha}^0}$).
\end{The}

\proo The result has been proved for every finite ordinal (integer) $\geq 1$ in \cite{fina}.
Let then $\alpha$ be a countable infinite ordinal. 
The above defined $\om$-CFL $ F(X,Y)$ is either a ${\bf \Pi^0_1}$-subset or a 
${\bf \Si^1_1}$-complete subset of $\{a, b\}^\om$. In the first case it is in the class 
${\bf \Si_{\alpha}^0}$ ( respectively ${\bf \Pi_{\alpha}^0}$) and in the second case it is not 
a Borel set. But one cannot decide which case holds. \ep 

\section{$\om$-powers of finitary languages}

\noi  We study in this section $\om$-powers of finitary languages, i.e. \ol s in the form 
$V^\om$ where $V$ is a finitary language. $\om$-powers of finitary languages are always 
analytic sets because whenever $V$ is finite, $V^\om$ is an \orl~ and then it is a boolean 
combination of ${\bf \Si^0_2}$-sets and whenever $V$ is 
countably infinite, one can fix an enumeration of $V$ and 
obtain $V^\om$ as a continuous image of $\om^\om$  
(the set of infinite sequences of integers $\geq 0$), 
\cite{sim}.
 
    Niwinski asked in \cite{niw} for an example of finitary language 
$W$ such that $W^\om$ is an analytic but non Borel set.
\nl From the results of preceding section, we can easily find an example of a context 
free language 
$W$ such that $W^\om$ is not a Borel set. 

    Consider the construction of the \ol~ $C$ from the \ol~ $B\subseteq \Si^\om$ in 
the proof of Theorem \ref{cfnotbor}. As stated above, if  
$g$ is  the substitution $\Si\ra P((\Si \cup \{A\})^\star)$ defined by
$a \ra a.D$ where 
$$D=\{ u.A.v ~\mid ~ u, v \in \Sis ~and~ ( |v|=2|u|)~~ or ~~( |v|=2|u|+1)~ \}$$

 \noi then $D$ is a context free language over the alphabet $(\Si \cup \{A\})$ and 
$g(B)=C$ holds. 

    Assume now that $B$ is an $\om$-power in the form $V^\om$. Then 
$g(B)=(g(V))^\om$ is also an $\om$-power.

    Let then $\Si=\{0,1\}$ be an alphabet containing two letters $0$ and $1$ and 
$W=0^\star.1$.
Then $W^\om=(0^\star.1)^\om$ is the set of $\om$-words over the alphabet $\Si$ which 
contain infinitely many occcurrences of the letter $1$.
It is a well known  example of an  \orl~ which is a ${\bf \Pi^0_2}$-complete subset 
of $\Si^\om$.

\noi Thus the language $g(W)$ is a finitary context free language such that 
$(g(W))^\om$ is an analytic but non Borel set. 
\nl  This language $g(W)$ is in fact a one counter 
language. 

    This gives an answer to Niwinski's question and additional answer to 
questions of Simonnet  who asked in \cite{sim} for the topological complexity 
of the $\om$-powers of context free languages. 

\section{Arithmetical properties}

We are going to deduce from the previous  proofs some new results 
about $\om$-context free languages  and the Arithmetical hierarchy. 
  We recall first the definition of the  Arithmetical hierarchy of  \ol s, \cite{sta}.

    Let $X$ be a finite alphabet. An \ol~ $L\subseteq X^\om$  belongs to the class 
$\Si_n$ if and only if there exists a recursive relation 
$R_L\subseteq (\mathbb{N})^{n-1}\times X^\star$  such that
$$L = \{\sigma \in X^\om ~\mid~\exists a_1 \ldots Q_na_n  \quad (a_1, \ldots , a_{n-1}, 
\sigma[a_n+1])\in R_L \}$$

\noi where $Q_i$ is one of the quantifiers $\fa$ or $\exists$ 
(not necessarily in an alternating order). An \ol~ $L\subseteq X^\om$  belongs to the class 
$\Pi_n$ if and only if its complement $X^\om - L$  belongs to the class 
$\Si_n$.  
\nl The inclusion relations that hold  between the classes $\Si_n$ and $\Pi_n$ are 
the same as for the corresponding classes of the Borel hierarchy.

\begin{Pro}[\cite{sta}]
\begin{enumerate}
\ite[a)] $\Si_n\cup \Pi_n \subsetneq  \Si_{n +1}\cap \Pi_{n +1}$, for each integer $n\geq 1$.
\ite[b)] A set $W\subseteq X^\om$ is in the class $\Si_n$ if and only if its 
complement $W^-$ is in the class $\Pi_n$. 
\ite[c)] $\Si_n - \Pi_n \neq \emptyset$ and $\Pi_n - \Si_n \neq \emptyset$ hold 
 for each integer $n\geq 1$.
\end{enumerate}
\end{Pro}

\noi The classes $\Si_n$ and $\Pi_n$ are strictly included in the respective classes 
${\bf \Si_n^0}$ and ${\bf \Si_n^0}$ of the Borel hierarchy:

\begin{The}[\cite{sta}]\label{arinbo}
For each integer $n\geq 1$, $\Si_n \subsetneq {\bf \Si_n^0}$ and 
$\Pi_n \subsetneq {\bf \Pi_n^0}$.
\end{The}

\noi Recall now  preceding results of \cite{fina}:

\begin{The}
Let n be an integer $\geq 1$. Then it is undecidable whether an effectively given $\om$-CFL
is in the class $ \Si_n$ ( respectively $ \Pi_n$).
\end{The}

\noi As in the case of the Borel hierarchy, projections of arithmetical sets 
(of the second $\Pi$-class) lead 
beyond the Arithmetical hierarchy, to the Analytical hierarchy of \ol s. The first class 
of this hierarchy is the class $\Si^1_1$. 
An \ol~ $L\subseteq X^\om$  belongs to the class 
$\Si_1^1$ if and only if there exists a recursive relation 
$R_L\subseteq (\mathbb{N})\times \{0, 1\}^\star \times X^\star$  such that:

$$L = \{\sigma \in X^\om ~\mid~\exists \tau (\tau\in \{0, 1\}^\om \wedge \fa n \exists m 
 ( (n, \tau[m], \sigma[m]) \in R_L )) \}$$

\noi Then an \ol~ $L\subseteq X^\om$  is in the class $\Si_1^1$ iff it is the projection 
of an \ol~ over the alphabet $X\times \{0, 1\}$ which is in the class $\Pi_2$ of the 
arithmetical hierarchy. 

    It turned out that an \ol~ $L\subseteq X^\om$ is in the class $\Si_1^1$
iff it is accepted by a non deterministic Turing machine (reading $\om$-words)
with a Muller acceptance condition \cite{sta}.
This class is denoted $NT(inf, =)$ (where $(inf, =)$ indicates the Muller condition) 
in \cite{sta} and also called the class of recursive 
\ol s $REK_\om$. 
\footnote{In another presentation, as in \cite{rog}, the recursive \ol s are those which are in 
the intersection $\Si_1\cap \Pi_1$, see also  \cite{lt}.}

    With the above definitions, one  can state the following:

\begin{The}[\cite{sta}]
 The class $CFL_\om$ is strictly included into the class $REK_\om$ of recursive \ol s.
\end{The}

    A natural question arises: are there $\om$-CFL  which are 
in the class $\Si_1^1$ but in not any class of the arithmetical hierarchy?
The answer can be easily derived from the preceding corresponding results about the 
Borel Hierarchy.

\begin{The}
There exist some context free  \ol s in $\Si_1^1 - \bigcup_{n\geq 1} \Si_n$.
\end{The}

\proo It follows from Theorems \ref{cfnotbor} and \ref{arinbo}.  \ep 

    We now obtain a recursive analogue to Theorem \ref{cfindbor}:

\begin{The}
Let $\Si$ be an alphabet containing at least two letters.
It is undecidable, for an effectively given context free   \ol~  $B$ to determine whether 
$B$  is in $\Si_1^1 - \bigcup_{n\geq 1} \Si_n$.
\end{The}

\proo  Recall that we had found (see proof of Theorem \ref{cfindbor})
a family of context free   \ol s $D(X,Y)$ 
over the alphabet 
$\Ga=\{a, b, c, \tla , d, A\}$ such that $D(X,Y)$ 
is either a ${\bf \Si^1_1}$-complete 
subset of $\Ga^\om$, or equal to $\Ga^\om$.

    Whenever $D(X,Y)$ is ${\bf \Si^1_1}$-complete, it is not in 
$\bigcup_{n\geq 1} \Si_n$ because each arithmetical class $\Si_n$ (respectively $\Pi_n$)
is included in the Borel class ${\bf \Si^0_n}$ (respectively ${\bf \Pi^0_n}$).
 
    Whenever $D(X,Y)$ is equal to $\Ga^\om$, $D(X,Y)$ 
  is in the class $\Si_1$ because of the characterization of \ol s in 
$\Si_1$ \cite{sta}: an \ol~ $L\subseteq X^\om$  belongs to the class 
$\Si_1$ if and only if there exists a recursive finitary language $W\subseteq X^\star$ 
such that $L=W.X^\om$.

    But we had proved that one cannot decide which of these two cases holds, hence the result
is proved for the alphabet $\Ga$.
(And we can use similar methods as in the proof of Theorem \ref{cfindbor} to obtain the result 
for an alphabet of cardinal $\geq 2$).   \ep

    Considering Turing machines, we get the following:

\begin{The} 
It is undecidable to determine whether the complement of an effectively given $\om$-CFL
is accepted by a non deterministic Turing machine with B\"uchi (respectively 
Muller) acceptance 
condition. 
\end{The}

\proo 
As in the preceding proof consider the 
family of context free   \ol s $D(X,Y)$ 
over the alphabet 
$\Ga=\{a, b, c, \tla , d, A\}$ such that $D(X,Y)$ 
is either a ${\bf \Si^1_1}$-complete 
subset of $\Ga^\om$, or equal to $\Ga^\om$.

    Whenever $D(X,Y)$ is ${\bf \Si^1_1}$-complete,  its complement 
is ${\bf \Pi^1_1}$-complete thus it is not a  ${\bf \Si^1_1}$ set (because a set 
which is both ${\bf \Si^1_1}$ {\bf and} ${\bf \Pi^1_1}$ is a Borel set) and therefore 
it is not a $\Si_1^1$-set (because the class $\Si_1^1$ is included in the class ${\bf \Si^1_1}$) 
then it is not accepted by any Turing machine with 
B\"uchi (respectively Muller) acceptance 
condition. 

    In the other case  $D(X,Y)$ is equal to $\Ga^\om$, then its complement is the emptyset 
and it is accepted by a Turing machine with 
B\"uchi (respectively Muller) acceptance 
condition. 

    But we had proved that one cannot decide which of these two cases holds, hence the result
is proved for the alphabet $\Ga$.
(And we can use similar methods as in the proof of Theorem \ref{cfindbor} to obtain the result 
for an alphabet of cardinal $\geq 2$).   \ep

    In fact this result can be extended to other \de machines.
Consider {\bf X}-automata as defined in \cite{eh} which are automata equipped with a storage 
type {\bf X}. 

\begin{The} Let {\bf X} be  a storage type  as defined in \cite{eh}. Then 
it is undecidable to determine whether the complement of an effectively given $\om$-CFL
is accepted by a non deterministic {\bf X}-automaton with B\"uchi (respectively 
Muller) acceptance 
condition. 
\end{The}

\proo It is similar to the previous one because every {\bf X}-automaton  
is less expressive than a Turing machine hence it cannot accept any 
${\bf \Pi^1_1}$-complete set. And conversely $\Ga^\om$ is accepted by every 
{\bf X}-automaton.   \ep 

\section{Context free languages of infinite trees}

The theory of automata reading infinite words have been extended 
to automata reading infinite binary trees labelled in a finite alphabet, i.e. trees in 
a space $T_\Si^\om$ where $\Si$ is a finite alphabet (and one may also consider infinite 
$k$-ary trees labelled in $\Si$ but we shall restrict ourselves here to binary trees), see 
\cite{tho}\cite{tho96}\cite{lt}\cite{sim} for many results and references.

    It is known that regular languages of infinite binary trees exhaust the hierarchy 
of Borel sets of finite rank as shown by Skurczynski \cite{sku}.
Niwinski proved that there exist some regular set of trees which are non Borel sets, 
 \cite{niw85}.

    Some regular sets of trees  are 
${\bf \Si^1_1}$-complete, as $Path(B)$ where $B$ is any ${\bf \Pi^0_2}$-complete
 regular subset of $\Si^\om$. 
$Path(B)$ (defined in the proof of Theorem \ref{cfnotbor}) is accepted by a non deterministic 
tree automaton which guesses a branch of a tree (using the non determinism) and then simulates 
a finite automaton on the path associated with this branch. 
\nl One can also define, for each \ol~ $B\subseteq \Si^\om$, the following sets of trees.
\nl Let  $$\fa - Path(B)$$
\noi be the set of trees $t$ in $T_\Si^\om$ such that every 
path of $t$ is in $B$, and let 
 $$Left-Path(B)$$  
\noi be the set of trees $t$ in $T_\Si^\om$ such that 
the leftmost path of $t$ is in $B$ (the nodes of the leftmost branch are  the 
words of $\{l, r\}^\star$ which are in the form $l^n$ for an integer $n\geq 0$).

    It is then well-known that whenever $B\subseteq \Si^\om$ is an \orl~ , the sets 
$\fa - Path(B)$ and $Left-Path(B)$  are regular sets of trees. Then if $B$ is a 
${\bf \Pi^0_2}$-complete subset of $\Si^\om$ it holds that:
$$\fa - Path(B^-) =  T_\Si^\om - (Path(B))$$
hence $\fa - Path(B^-)$ is a  ${\bf \Pi^1_1}$-complete  subset of $T_\Si^\om$.

    The Theorem of complementation of Rabin implies that every regular 
set of trees is in ${\bf \Si^1_2} \cap {\bf \Pi^1_2}$, and it has been shown 
that there exist  regular sets of trees which are not in 
${\bf \Si^1_1} \cup {\bf \Pi^1_1}$, see  \cite{lt} for a view of a  hierarchy of 
regular sets of trees. 

    As  finite automata have been extended to (top-down)  automata
on infinite trees, pushdown automata have been extended to (top-down)  pushdown automata
on infinite trees by Saoudi \cite{sao}. Denote, as in \cite{sao},  $CF_3$ the family 
of languages of infinite (binary) trees accepted by (top-down)  
pushdown automata with Muller acceptance condition. 

    It is easy to see from the definition of these automata that, as in the case of  
 tree automata, if $B$ is an $\om$-CFL, then the sets of trees 
$Path(B)$  and $Left-Path(B)$  are accepted by tree
pushdown automata. Then we can extend our preceding undecidability results 
of Theorems \ref{cfindbor} and \ref{indbor2}.

\begin{The}\label{indtree}
\begin{enumerate}
\ite[(a)] Let $\alpha$ be a countable ordinal $\geq 1$. 
Then it is undecidable to determine whether an effectively given language in  $CF_3$
is in the Borel class ${\bf \Si_{\alpha}^0}$ ( respectively ${\bf \Pi_{\alpha}^0}$).
\ite[(b)] It is undecidable to determine whether an effectively given language in  $CF_3$ 
is a Borel set.
\ite[(c)] It is undecidable to determine whether an effectively given language in  $CF_3$ 
is in the class ${\bf \Pi_1^1}$ .
\ite[(d)] It is undecidable to determine whether an effectively given language in  $CF_3$ 
is a ${\bf \Si_1^1}$  but non Borel set.
\end{enumerate}

\end{The}

\proo  The proofs  are easily derived from the proof of 
Theorem \ref{cfindbor}. Recall we had got a family of omega context free languages  $D(X,Y)$ 
over the  alphabet $\Si_A$ such that: either $D(X,Y)=(\Si_A)^\om$, or   $D(X,Y)$ 
is a ${\bf \Si_1^1}$-complete subset of $(\Si_A)^\om$. But one cannot decide which case holds. 
\nl  It is easy to see that $Left-Path(D(X,Y))$  has the same
 topological complexity  as 
the \ol~  $D(X,Y)$. 
\nl Indeed let $f$ be the function: $(\Si_A)^\om \ra T_{\Si_A}^\om $ defined by 
$f(\sigma)=t_\sigma$ where $t_\sigma$ is the tree in $T_{\Si_A}^\om $ with $\sigma$ as 
leftmost path and the letter $A$ labelling the other nodes. Then $f$ 
is continuous and $f^{-1}(Left-Path(D(X,Y)))=D(X,Y)$.  
 Assume first that  $D(X,Y)$ 
is a ${\bf \Si_1^1}$-complete subset of $(\Si_A)^\om$, then $Left-Path(D(X,Y))$  is also 
at least 
${\bf \Si_1^1}$-complete and not a  Borel set.
\nl Now let $j$ be the function $T_{\Si_A}^\om \ra (\Si_A)^\om $ defined by: 
$j(t)$ is the leftmost path of the tree $t$. Then $j$ is a continuous function and 
$j^{-1}(D(X,Y))=Left-Path(D(X,Y))$. Hence when $D(X,Y)$ 
is a ${\bf \Si_1^1}$-complete subset of $(\Si_A)^\om$, $Left-Path(D(X,Y))$  is a 
${\bf \Si_1^1}$-set because the class ${\bf \Si_1^1}$ is closed under inverse of continuous 
functions. Thus $Left-Path(D(X,Y))$  is a ${\bf \Si_1^1}$-complete subset of 
$T_{\Si_A}^\om$ and not 
 a  ${\bf \Pi_1^1}$-set. 
\nl In the other 
case $D(X,Y)=(\Si_A)^\om$ and 
$Left-Path(D(X,Y)) = T_{\Si_A}^\om $  then $Left-Path(D(X,Y))$  is in every Borel class 
and also in the class ${\bf \Pi_1^1}$. 
But one cannot decide which case holds. This proves $(a)$, $(b)$, $(c)$ and $(d)$. \ep

\section{Concluding remarks and further work}

\noi We have proved in \cite{fina} that the class of $\om$-CFL exhausts the finite ranks of the 
 Borel hierarchy and in this paper (Theorem \ref{cfnotbor}) that there 
exist some analytic but non Borel $\om$-CFL. 
\nl The question to know whether there exist some $\om$-CFL which are Borel sets of 
infinite rank is still open.

There exists a refinement of the  Borel hierarchy which is called the Wadge 
hierarchy of Borel sets. We proved in \cite{finb} that the length of the Wadge hierarchy of 
$\om$-CFL is an ordinal greater than or equal to the Cantor ordinal $\varepsilon_0$.
And it remains to find the exact length of the  Wadge 
hierarchy of Borel $\om$-CFL.

Mention that on the other side, the Wadge hierarchy of \de $\om$-CFL has been determined.  
its length is the ordinal $\om^{(\om^2)}$. It has been recently studied 
in \cite{dfr} \cite{dupcf} \cite{fidet}.

    {\bf  Acknowledgments.} We have  
 previously proved  the existence of analytic but non Borel sets in 
 another class of \ol s, the class of locally finite \ol s \cite{omloc}. 
We are indebted  
to Jean-Pierre Ressayre who suggested the way to adapt the original proof to the  
context free case.

\end{document}